\documentstyle[eqsecnum,twocolumn,aps,epsf]{revtex}
\newcommand{\eps}{\varepsilon}
\newcommand{\teps}{\tilde{\varepsilon}}
\newcommand{\vphi}{\varphi}
\begin{document}
\draft
\preprint{Condmat/xxxxxxx}

\title{Local Inhomogeneity Effects on Nucleation Process \\
in a High External Bias
\thanks{Submitted to Phys. Rev. B.}
}
\author{Takeo Kato\thanks{E-mail address: kato@a-phys.eng.osaka-cu.ac.jp}}
\address{
Department of Applied Physics, Osaka City University, Sumiyoshi-ku,
Osaka 558-8585, Japan}
\date{\today}
\maketitle
\begin{abstract}
Quantum nucleation processes in the presence of local moderate
inhomogeneities are studied theoretically at high biases.
The quantum nucleation rate Gamma is calculated for one-dimensional
systems in a form $\Gamma = A e^(-B/\hbar)$ by using the `bounce' method.
The bias-dependence of the exponent $B$ is shown to be changed by 
inhomogeneities. This change is explained by the reduction of the 
effective spatial dimension of the system. By studying the system-size 
dependence of the prefactor $A$, the condition for the appearance of 
inhomogeneity effects is evaluated. Nucleation rates in thermal 
activation regimes are also calculated, and compared with quantum 
tunneling regimes. For higher-dimensional systems, it is shown that 
the local approximation of inhomogeneity does not hold, and that 
spatial profiles of inhomogeneity become important.
\end{abstract}

\narrowtext

\section{Introduction}
\label{sec:introduction}

Nucleation process is one of the most universal phenomena
found in various areas in physics from cosmology to condensed
matter physics providing the mechanism 
for the onset of first-order transitions \cite{Haenggi90}.
Besides classical nucleation due to thermal fluctuations,
nucleation due to quantum tunneling 
has attracted interest for many years as macroscopic quantum
phenomena. Although quantum nucleation has been studied theoretically
in the pioneering issue \cite{Lifshitz72}, it is just recently that 
observation of quantum nucleation has become possible owing to
the progress in experimental techniques.
At present, quantum nucleation has been observed experimentally
in low-temperature condensed-matter systems: 
${\rm }^4{\rm He}$-${\rm }^3{\rm He}$ liquid solutions 
\cite{Satoh91,Satoh92},
cavitation in ${\rm }^4{\rm He}$ liquid \cite{Balibar95}, nucleation of
${\rm }^4{\rm He}$ solids \cite{Ruutu96}. 
Nucleation of magnetic domain in thin films
has also been discussed theoretically 
\cite{Caldeira88,Chudnovsky88,Chudnovsky98}.
Quantum creation of a kink-antikink pair can be regarded as 
one-dimensional nucleation, and has been
studied both experimentally and theoretically
in dislocation motion in solids \cite{Takeuchi00,Petukhov98},
and sliding of charge density waves (CDWs)
\cite{Maki78,Hida84,Nakaya86,Duan93,Maki95,Hatakenaka98},
though in the latter system interpretations of the experimental 
results are not settled \cite{Zotov93,Zotov94,Zotov97}. 
I believe that long Josephson 
junctions \cite{Davidson85,Ustinov92,Castellano96,Kato96,Ustinov99}
are also suitable to observe quantum nucleation of a soliton pair.

In many theoretical issues, nucleation processes have been 
restricted to the situation that a stable phase is 
formed in the otherwise {\it homogeneous} background of 
an unstable phase. In several literatures, 
the nucleation rates in the presence of strong 
inhomogeneities have also been studied by using a
single-variable model \cite{effective}. 
In such studies, direct evaluation of the impurity
strength from the nucleation rate is difficult, 
because the parameters of the 
single-variable model cannot be related clearly
to the impurity strength. In this paper, we study
{\it moderate} inhomogeneities, which can be
controlled by an external parameter 
from the nearly-homogeneous region
to the strongly-inhomogeneous region.
This crossover region has not been
studied in detail, and provides useful information
of the impurity strength, because the nucleation
rates are very sensitive to the impurity strength 
in this region.

In this paper, we study how the nucleation process 
changes in the presence of local moderate inhomogeneity.
As a starting point, a highly-biased region is investigated
where the potential can be expressed by a cubic polynomial.
The nucleation rate is calculated based on the `bounce' method 
\cite{Coleman77,Callan77,Coleman79} for quantum tunneling regimes,
and on Kramers' law for thermal activation 
regimes \cite{Haenggi90,Kramers40}.
We concentrate on nucleation processes in one-dimensional
systems which correspond to the kink-antikink nucleation.
We show that extension to higher dimensions
is not easy since the nucleation rates are 
affected by details of local inhomogeneity profiles
such as an impurity size.

The nucleation rate is expressed in the form 
$\Gamma = A \exp (-B/\hbar)$,
where $A$ and $B$ are called as a prefactor and an exponent, respectively.
The exponent $B$ depends on the bias $f$ as
\begin{equation}
B \propto (f_c - f)^{\gamma}, 
\end{equation}
where $f_c$ is a classical threshold bias at which the potential
barrier disappears. It should be stressed
that the value of $\gamma$ is modified by inhomogeneities.
The modification of $\gamma$
has first been discussed in the quantum sliding of CDWs 
by Yumoto {\it et.al} \cite{Yumoto99,Yumoto00a,Yumoto00b}.
They have calculated the nucleation rate by reducing
the model to a single-variable problem with use of
path integrals, and have reported that
the value of $\gamma$ increases as the impurity strength
is enlarged. This method, however, is so complicated that
it is difficult to discuss the origin of the change of $\gamma$
clearly, and also to examine the validity of the approximation
adopted there. In this paper, we study inhomogeneity
effects without any reduction to a single-variable problem.
It is claimed that our results are explained by
the `dimensionality' of nucleation, which has been pointed out
by the author in ref. \cite{Kato00}. 

We consider one isolated impurity in a system
with the size $L$. The results obtained
in this paper are also applicable to systems 
with the dilute impurities
by taking the average impurity distance as $L$.
In the limit $L\rightarrow \infty$, 
nucleation occurs dominantly in homogeneous 
regions of the samples, and no impurity effects appear. 
The inhomogeneity effects
appear only when $L$ is below a crossover value $L_{\rm cr}$.
Within the bounce method, $L_{\rm cr}$ is shown
to be obtained by calculating the prefactor $A$.
Here, we should note that the validity of the
bounce method is not guaranteed generally for the many-body problems.
We also discuss the validity of the bounce method 
when it is applied to the nucleation problem in the presence 
of the impurities.

This paper is organized as follows.
The model Hamiltonian is given in Sec. \ref{sec:model}.
The quantum tunneling rate is formulated and calculated
for one-dimensional systems in Sec. \ref{sec:quantum}. 
We study the thermal regime in Sec. \ref{sec:thermal} briefly, and
the extension to higher-dimensional systems
and justification of the bounce method are discussed
in Sec. \ref{sec:discussion}.
Finally, results are summarized in Sec. \ref{sec:summary}.

\section{Model}
\label{sec:model}

In this paper, nucleation rates are calculated based on the equation
of motion
\begin{equation}
\phi_{tt} - \nabla^2 \phi + \frac{\partial V}{\partial \phi} = 0.
\label{eq:KGeq0}
\end{equation}
Here, $\phi(\vec{x},t)$ is a $(d+1)$-dimensional field, and
$x$ and $t$ are scaled by the characteristic
length and time, respectively.
Dissipation is assumed to be weak enough, 
but not extremely weak so that the system is in a thermal 
equilibrium in a metastable well. 
We concentrate on one-dimensional systems described
by the equation
\begin{equation}
\phi_{tt} - \phi_{xx} + \frac{\partial V}{\partial \phi} = 0,
\label{eq:KGeq}
\end{equation}
and the higher-dimensional systems are discussed only
in Sec. \ref{sec:discussion}.
The potential energy is assumed to consist of two parts
as $V(\phi,x)=V_0(\phi)+V_{\rm imp}(\phi,x)$, where
$V_0(\phi)$ is a homogeneous part and
$V_{\rm imp}(\phi,x)$ describes an inhomogeneity.

As for the homogeneous part $V_0(\phi)$, 
we assume: (1) the potential has at least one metastable 
point $\phi_0$; (2) the energy barrier in $V_0(\phi)$
is controlled by an external parameter like an external field.
One typical example for $V_0(\phi)$ is a tilted cosine potential 
\begin{equation}
V_0(\phi) = (1-\cos \phi) - f\phi.
\label{eq:pot1}
\end{equation}
Another example is the $\phi^4$-model with a bias
\begin{equation}
V_0(\phi) = - \frac{\phi^2}{2} + \frac{\phi^4}{4!} - f\phi. 
\label{eq:pot2}
\end{equation}
The type in (\ref{eq:pot1}) has been used for long Josephson junctions,
dislocation in solids, and CDW systems, while the type
in (\ref{eq:pot2}) describes an effective model for 
nucleation of a stable phase from an unstable background.

In this paper, the characteristic action of the systems is 
assumed to be large as compared with
the Planck constant. In such situations, the quantum nucleation
appears only when the bias $f$ is controlled near 
the classical threshold $f_c$ at which the metastable 
state of the potential $V_0(\phi)$ disappears. 
Note that $f_{\rm c,0}=1$ for the tilted cosine
potential and $f_{\rm c,0}=2\sqrt{2}/3$ for the $\phi^4$-model
in the absence of impurities. 
Near $f = f_{\rm c,0}$, the potentials in (\ref{eq:pot1}) and 
(\ref{eq:pot2}) form can be approximated commonly by
quadratic-plus-cubic polynomials
around a metastable point $\phi=\phi_0$ as
\begin{eqnarray}
V_0(\phi) - V_0(\phi_0) &\simeq& 
\frac{a (f_{\rm c,0}-f)^{1/2}}{2} 
(\phi-\phi_0)^2 \nonumber \\
&-& \frac{b}{6} (\phi-\phi_0)^3.
\label{homopot}
\end{eqnarray}
Here, $a = O(1)$ and $b=O(1)$.
To be explicit, $a =2^{1/2}$, $b = 1$
for the tilted cosine potential (\ref{eq:pot1}),
while $a=2^{3/4}$, $b = 2^{1/2}$ for the $\phi^4$-model 
(\ref{eq:pot2}).

\begin{figure}[tb]
\hfil
\epsfxsize=8cm
\epsfbox{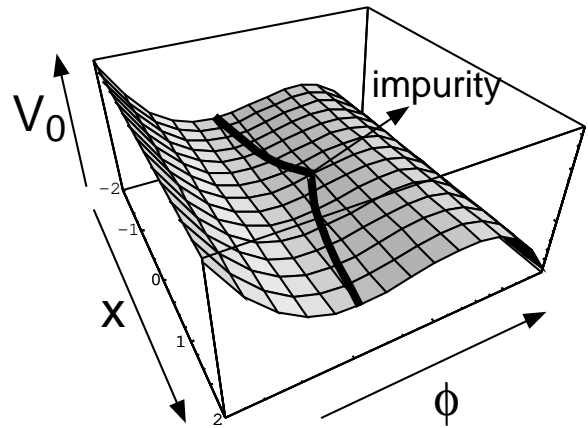}
\hfil
\caption{String motion in the potential $V_0(\phi)$.
The impurity effect is described by a local force for
this string at $x=0$.}
\label{fig:pot}
\end{figure}

The equation of motion (\ref{eq:KGeq})
describes dynamics of a string in the potential $V(\phi)$ shown in 
Fig. \ref{fig:pot}. When the string is initially located
in a metastable well, it stays there for a while, and then
begins to move towards lower energy states.
If the system size is large enough,
the transition from metastable states to moving states
occurs through local deformation of the string accompanied
with the creation of one kink-antikink pair.
This transition can be regarded as `nucleation'
when the metastable(stable) state is related to the 
metastable(stable) phase. 

The local inhomogeneity is introduced by $V_{\rm imp}(\phi)$ as
\begin{equation}
V_{\rm imp}(\phi,x) = \delta(x) h(\phi),
\label{Vimp}
\end{equation}
where $\delta(x)$ is a delta function, and $h(\phi)$ is a
function defined around the metastable point. 
This type of impurity potentials has been studied theoretically 
in CDW systems \cite{Yumoto99,Yumoto00a,Yumoto00b} 
and long Josephson junctions \cite{McLaughlin78,Kivshar89}.
The potential form (\ref{Vimp}) can be obtained not only for
impurities but also for inhomogeneities of external forces,
modulation of barrier heights, and edge effects. (Details
are summarized in Appendix \ref{app}.)
At high biases, since the impurity effect is 
determined by a small modulation of $\phi$ around
the metastable point $\phi_0$,
the inhomogeneity potential can be approximated as
\begin{equation}
V_{\rm imp}(\phi,x) \sim - \eps \delta(x) (\phi-\phi_0).
\label{impurity}
\end{equation}
Here, $\eps$ describes the impurity strength.
This local potential can be regarded as a local force on
a string at $x=0$ as shown in Fig. \ref{fig:pot}.
For $\eps > 0$ ($\eps < 0$), 
the string is attracted towards the positive(negative)
direction of $\phi$. As a result,
the nucleation is enhanced around $x=0$ for $\eps > 0$, 
and suppressed for $\eps < 0$.

The appearance of the impurity effect is explained intuitively
as follows. The nucleation rate $\Gamma$ is expected to 
consist of two parts as $\Gamma = \Gamma_{\rm bulk} + \Gamma_{\rm imp}$.
Here, $\Gamma_{\rm bulk}$ is the nucleation rate 
in the homogeneous region far from an impurity, and $\Gamma_{\rm imp}$ 
is a contribution near the impurity. Using the system size $L$,
the bulk part is estimated as $\Gamma_{\rm bulk} 
\sim \Gamma_0 L$, where $\Gamma_0$ is the nucleation rate per
unit length in the absence of the impurities. The impurity part
$\Gamma_{\rm imp}$ is independent of $L$, and 
is estimated as $\Gamma_{\rm imp}\sim\Gamma_0 \exp(\Delta U/\hbar
\omega_0)$, where $\Delta U$ is the energy gain at the impurity
site, and $\omega_0$ is an attempt frequency around the metastable
state. If the system size $L$ is large
enough, the bulk part $\Gamma_{\rm bulk}$ dominates the impurity 
part $\Gamma_{\rm imp}$, and no impurity effects appear. 
However, by controlling $\Delta U$, $\Gamma_{\rm imp}$ 
can dominate $\Gamma_{\rm bulk}$. The
inequality $\Gamma_{\rm imp} \gg \Gamma_{\rm bulk}$ leads
\begin{equation}
\Delta U \gg \hbar \omega_0 \log L.
\label{condition}
\end{equation}
Since the condition (\ref{condition}) shows of a weak
system-size dependence of the condition 
(\ref{condition}) through the logarithmic function of $L$,
it is expected to be feasible in experimental systems.
The inequality (\ref{condition}), which can be derived more 
accurately in Sec. \ref{sec:prefactor}, will be assumed in the next
section.

In this paper, we further assume
$\eps > 0$ in (\ref{impurity}), 
because $\Gamma_{\rm imp}$ never dominates the bulk part
$\Gamma_{\rm bulk}$ in case of $\eps < 0$.

\section{QUANTUM TUNNELING REGIME}
\label{sec:quantum}

\subsection{Formulation and Scaling Analysis}

The quantum nucleation rate in the presence of the impurity
is formulated by a semiclassical method as follows.
The partition function of the system
is expressed in terms of path integrals as
\begin{eqnarray}
Z &=& \int {\cal D}\phi(x,\tau) \exp \left( 
- \frac{S[\phi(x,\tau)]}{\hbar}\right), 
\label{partition1} \\
S[\phi(x,\tau)] &=& 
\int_{-L/2}^{L/2} {\rm d}x \int_0^{\hbar \beta} 
{\rm d}\tau \left[
\frac{1}{2}\phi_x^2 + \frac{1}{2} \phi_{\tau}^2 
\right. \nonumber \\ &+& \left. V_0(\phi) + 
V_{\rm imp}(x,\phi) \right] ,
\label{partition2}
\end{eqnarray}
where $\hbar$ is a dimensionless Planck constant
normalized by the characteristic frequency, length,
and energy of the system, $\tau=i t$ is an imaginary
time, and $\beta = 1/k_{\rm B}T$ is an inverse temperature. 
The potential forms
are given by (\ref{homopot}) and (\ref{impurity}).
By rescaling the variables as
\begin{eqnarray}
x &=& x' \left(a \sqrt{f_{\rm c,0} - f}\right)^{-1/2}, 
\label{cspace} \\
\tau &=& \tau' \left(a \sqrt{f_{\rm c,0} - f}\right)^{-1/2}, 
\label{ctmp} \\
\phi &=& \frac{a \sqrt{f_{\rm c,0} - f}}{b} \vphi + \phi_0,
\end{eqnarray}
the action is rewritten as
\begin{eqnarray}
& & S[\vphi(x,\tau)] = \frac{a^2 (f_{\rm c,0} - f)}{b^2}
\int_{-\tilde{L}/2}^{\tilde{L}/2} 
{\rm d}x' \int_0^{\hbar \tilde{\beta}} 
{\rm d}\tau' \nonumber \\
& & \hspace{10mm} \times \left[
\frac{1}{2}\vphi_x^2 + \frac{1}{2} \vphi_{\tau}^2 
+ \frac{1}{2} \vphi^2 - \frac{1}{3!} \vphi^3 - \tilde{\eps}
\delta(x') \vphi \right].
\label{action2}
\end{eqnarray}
Here, $\tilde{L}=L a^{1/2}(f_{\rm c,0}-f)^{1/4}$ and
$\tilde{\beta} = \beta a^{1/2} (f_{\rm c,0}-f)^{1/4}$ is 
a scaled length and inverse temperature, respectively.
The impurity effect is described only through an effective
impurity strength $\tilde{\eps}$ defined by
\begin{equation}
\tilde{\eps} = \eps a^{-3/2} b (f_{\rm c,0} - f)^{-3/4}.
\end{equation}
This effective impurity strength depends on both the 
bare impurity strength $\eps$ and external bias $f$.
Even if $\eps$ is fixed, we can
control the strength of the impurity effect by
changing the external current; the impurity effect is 
enhanced by controlling the external current near the
threshold current $f_c$. 

The partition function given in 
(\ref{partition1}) and (\ref{partition2})
is evaluated by integrating the partition function around
stationary solutions up to second-order fluctuations.
The stationary solutions are determined by 
\begin{equation}
\frac{\delta S}{\delta \vphi} = 
- \vphi_{\tau' \tau'} - \vphi_{x'x'} + \vphi - \frac{\vphi^2}{2}
- \tilde{\eps} \delta(x') = 0.
\label{eq:bounceeq}
\end{equation}
There exist two types of solutions for this equation.
One is a stable solution $\vphi_0(x')$ independent of $\tau'$,
and the other is a `bounce' solution $\vphi_{\rm B}(x',\tau')$.
Since the system has a metastable state, 
the free energy $F = -\beta^{-1} \log Z$ 
gains an imaginary part. Then, the decay rates are given by
$\Gamma = 2 {\rm Im} F$. This formula, derived through
the `bounce' method, was first applied 
to the thermal activation regime \cite{Langer67,Langer69}, 
and extended to the quantum tunneling
regime \cite{Lifshitz72,Coleman77,Callan77,Coleman79}.
The nucleation rate is now written in the form
\begin{equation}
\Gamma = A \exp(-B/\hbar), 
\end{equation}
and the exponent $B$ and prefactor $A$ are given as
\begin{eqnarray}
B &=& S[\vphi_{\rm B}(x',\tau')] - S[\vphi_0(x')], 
\label{eq:exponent} \\
A &=& \frac{\bar{a}(T)}{\beta \hbar}
\prod_{i=1}^{\infty} \left(\frac{|\lambda_i^{\rm (B)}|}
{\lambda_i^{(0)}}\right)^{-1/2}.
\label{eq:prefactor}
\end{eqnarray}
Here, $\bar{a}(T)$ is a temperature-dependent factor,
and takes 1 for $T<T_0$, and $T_0/T$ for $T>T_0$ \cite{Affleck81},
and $\lambda_i^{\rm (B)}$s ($\lambda_i^{(0)}$s) are
the frequencies of eigenmodes around the bounce(stable) solution
determined by the equation
\begin{eqnarray}
& & (-\partial_{\tau' \tau'}-\partial_{x'x'}+1-\phi_{\rm B,0}(x',\tau'))
\psi^{\rm (B,0)}_i(x',\tau') \nonumber \\
& & \hspace{5mm}
= \lambda_i^{\rm (B,0)} \psi^{\rm (B,0)}_i(x',\tau').
\label{eq:schrodinger}
\end{eqnarray}
All the eigenvalues around the stable solution are positive
($\lambda^{(0)}_i \ge 1$).
The lowest eigenvalue $\lambda^{(B)}_1$ 
is negative due to the metastability of the bounce solution
$\vphi_{\rm B}$.
The second lowest eigenvalue $\lambda^{(B)}_2$
is always zero due to the translational symmetry
of $\vphi_{\rm B}$ in the $\tau$ direction. This mode
must be treated by the Fadeev-Popov method, and
the expression of the prefactor is modified as \cite{Callan77,Weiss93}
\begin{equation}
\frac{1}{\sqrt{\lambda_2^{(B)}}} \rightarrow 
\sqrt{\frac{B}{2\pi \hbar}}
\int_0^{\hbar \tilde{\beta}}d\tau_0' ,
\label{eq:improvement}
\end{equation}
where $\tau_0'$ denotes the center position of the bounce.
For the weak impurity, the third eigenvalue 
$\lambda_3^{(B)}$ also approaches zero due to the
translational symmetry of $\vphi_{\rm B}$ in the $x$-direction. 
We study this mode in Sec. \ref{sec:prefactor} in detail.

\subsection{Dimensional Crossover}
\label{sec:dimcross}

The bias dependence of the exponent $B$
can be discussed from the viewpoint of the `dimensional crossover'. 
This viewpoint has been discussed in several physical systems
such as CDWs \cite{Hatakenaka98}, and long Josephson 
junctions \cite{Ivlev87} in the context of the system-size dependences.
The application to impurity problems
has been discussed recently in ref. \cite{Kato00}. 
Here, we briefly summarize this viewpoint.

For the weak impurity $\tilde{\eps} \ll 1$,
the bias-dependence of the exponent $B$ is 
obtained as
\begin{equation}
B \propto (f_{\rm c,0}-f)^{\frac{3}{2}-\frac{d+1}{4}}.
\label{eq:cs1}
\end{equation}
Note that the exponent of $(f_{\rm c,0}-f)$ depends 
crucially on the spatial dimension $d$. The factor 
$3/2$ comes from the bias dependence of the 
barrier height $\Delta U \propto (f_{\rm c,0}-f)^{3/2}$,
while the factor $-(d+1)/4$ comes from the fact that
the typical size of the bounce solution in
the spatial(temporal) coordinates is given
by $(f_{\rm c,0}-f)^{-1/4}$ for each of original coordinates
($x$, $\tau$). (See (\ref{cspace}) and (\ref{ctmp}).)

The impurity effect can be explained by the change in the
dimension: the strong impurity changes
the spatial dimensionality $d$ to zero. 
The classical threshold bias is also modified from
$f_{\rm c,0}$ to $f_{\rm c}$.
As a result, the bias-dependence of $B$ becomes
\begin{equation}
B \propto (f_c-f)^{\frac{3}{2}-\frac{1}{4}} = (f_c-f)^{5/4}
\label{exp1}
\end{equation}
in the presence of strong impurities.
In this region, the nucleation process is well described
by a local deformation of the field, and is reduced
to a one-variable problem.

This viewpoint is valid in the one-dimensional case as shown 
in the subsequent sections.
In the $d\ge 2$ case, however, the viewpoint of the dimensional
crossover in this paper cannot be discussed,
since the impurity potential cannot be described by the delta function
in (\ref{impurity}), as will be discussed in Sec. \ref{sec:discussion}.

\subsection{Exponent}
\label{sec:exponent}

For one-dimensional systems, the exponent $B$ is calculated
in a form $B = \sigma \tilde{B}(\tilde{\eps})$. The factor
$\sigma = a^2 (f_c - f)/b^2$ describes the bias-dependence
in the absence of the impurities, while the scaled
exponent $\tilde{B}(\eps)$ is calculated as
\begin{eqnarray}
\tilde{B}(\tilde{\eps}) &=& \tilde{S}[\vphi_{\rm B}(x',\tau')]
- \tilde{S}[\vphi_0 (x')], \\
\tilde{S}[\vphi(x',\tau')] &=&
\int_{-\tilde{L}/2}^{\tilde{L}/2} dx' 
\int_0^{\hbar \tilde{\beta}} d\tau'
\left[ \frac{1}{2} \vphi_{x'}^2 + 
\frac12\vphi_{\tau'}^2 \right. \nonumber \\
&+& \left. \frac12 \vphi^2 - \frac16 \vphi^3
- \tilde{\eps} \delta (x') \vphi \right].
\label{eq:scaled}
\end{eqnarray}
The impurity effect appears only 
through $\tilde{B}(\tilde{\eps})$. 
In the homogeneous case, 
$\tilde{B}(\teps)$ is obtained as 
\begin{equation}
\tilde{B}(0) = 
\tilde{S}[\vphi_{\rm B}(x',\tau';\teps=0)] = 31.00,
\end{equation}
and the exponent behaves as $B = \tilde{B}(0) 
\sigma \propto (f_c - f)^1$.
As the effective impurity strength $\tilde{\eps}$ increases, 
$\tilde{B}(\tilde{\eps})$ is suppressed, and
the tunneling rate is enhanced. At a critical value 
$\tilde{\eps}_{\rm cl} = 4/\sqrt{3} \simeq 2.31$, the potential
barrier disappears, and the exponent $B$ is reduced to zero.
Note that before the exponent $B$ becomes zero, the bounce method becomes
invalid because it is justified only in the semiclassical
condition $B/\hbar \gg 1$.
However, the region where the bounce method is broken down is
narrow since the normalized Planck constant $\hbar$ is 
assumed to be small enough.

\begin{figure}[tb]
\hfil
\epsfxsize=8cm
\epsfbox{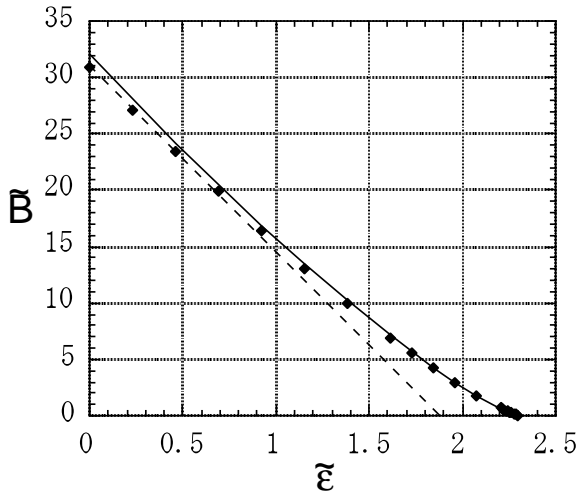}
\hfil
\caption{The scaled exponent $\tilde{B}(\teps)$
as a function of the effective impurity $\teps$. The numerical
result is shown by the square dots. Dashed and solid lines
corresponds to the analytical
results for the weak impurities (\ref{eq:weakexp})
and for the strong impurities (\ref{eq:strongfit}),
respectively.}
\label{fig:epsplot}
\end{figure}

We calculate the function $\tilde{B}(\tilde{\eps})$
by solving the equation (\ref{eq:bounceeq}) numerically.
The Newton method is used by dividing the $x'$-$\tau'$ 
($x'>0$, $\tau'>0$) space into $64\times 64$ cells. 
The result is shown in Fig. \ref{fig:epsplot} by square dots.

Analytical expressions of $\tilde{B}(\teps)$
can be obtained in the limiting cases
\cite{Kato00}. For the weak impurity ($\teps \ll 1$),
$\tilde{B}(\tilde{\eps})$ is
evaluated by using the bounce solution in
the homogeneous case $\teps = 0$ as
\begin{eqnarray}
\tilde{B}(\teps) &=& \tilde{S}_0
- \tilde{\eps} \int_{-\infty}^{\infty} 
d\tau' \vphi_{\rm B}(0,\tau';\tilde{\eps}=0) 
+ O(\tilde{\eps}^2) \nonumber \\
&=& 31.00 - 16.43 \tilde{\eps} + O(\tilde{\eps}^2).
\label{eq:weakexp}
\end{eqnarray}
This result is shown in Fig. \ref{fig:epsplot} by the dashed line. 

For the strong impurity $\tilde{\eps} \sim 
\tilde{\eps}_{\rm cl}$, the local modulation of the field becomes
relevant, and a one-mode approximation is effective.
We first evaluate the solutions of (\ref{eq:bounceeq}) 
for $\tilde{\eps} = \tilde{\eps}_{\rm cl}$. 
The bounce solution $\vphi_{\rm B}(x')$ agrees with the
stable solution $\vphi_0(x')$, and is obtained analytically as
\begin{equation}
\vphi_{\rm B}(x';\tilde{\eps}_{\rm cl}) 
= \frac{3}{\cosh^2((x'+a_{\rm cl})/2)},
\end{equation}
where $\tanh(a_{\rm cl}/2) = 1/\sqrt{3}$.
At the critical point $\teps = \teps_{\rm cl}$,
there appears the zero frequency mode ($\lambda^{(B)}_1 =0$)
around the solution $\vphi_{\rm B}(x;\teps_{\rm cl})$. 
The eigenfunction of this mode is obtained 
from (\ref{eq:schrodinger}) as 
\begin{equation}
\psi^{(B)}_1(x';\teps) = \frac{C \sinh((x'+a_{\rm cl})/2)}
{\cosh^3((x'+a_{\rm cl})/2)},
\end{equation}
where $C$ is a normalization factor determined from
$\int |\psi^{(B)}_1|^2 dx' = 1$, and is given by
\begin{equation}
C = \sqrt{\frac{135}{8(9-2\sqrt{3})}} \simeq 1.746.
\end{equation}
When $\teps$ is slightly smaller than 
$\teps_{\rm cl}$, this mode
describes slow dynamics of the system, and is related
to the tunneling process. 
Hence, we can approximate the bounce solution as
\begin{equation}
\vphi_{\rm B}(x',\tau';\teps) \simeq
\vphi_{\rm B}(x';\teps_{\rm cl})
+ X(\tau') \psi^{\rm (B)}_1 (x'),
\label{eq:tranc}
\end{equation}
where $X(\tau')$ is a tunneling variable denoting
the dynamics of the local deformation at the impurity site. 
By substituting (\ref{eq:tranc}) into the action (\ref{eq:scaled}),
we obtain the action in terms of $X(\tau')$ as
\begin{eqnarray}
\tilde{S}[X(\tau')] &=& \int_0^{\hbar \tilde{\beta}} d\tau'
\left[ \frac12 \left(\frac{dX}{d\tau'}\right)^2
+ V(X) \right] 
\label{aaa1} \\
V(X) &=& \frac{2\sqrt{3}}{9} (\teps_{\rm cl}-\teps) C X 
- \frac{4}{243} (C X)^3.
\label{aaa2}
\end{eqnarray}
This action corresponds to one-particle dynamics 
in a cubic potential $V(X)$. 
The potential barrier $\Delta U$
and frequency of small oscillation
around the metastable state $\omega_0$ are given as
\begin{eqnarray}
\Delta U &=& 2^{5/2} \times 3^{-5/4} (\teps_{\rm cl}-\teps)^{3/2} 
\label{eq:barrior} \\
\omega_0 &=& 2^{5/4} \times 3^{-11/8} C (\teps_{\rm cl}-\teps)^{1/4}
\label{eq:omega0}
\end{eqnarray}
By applying the standard bounce technique \cite{Coleman77,Coleman79}
to this potential, the function $\tilde{B}(\teps)$ is obtained 
\begin{equation}
\tilde{B}(\teps) = \frac{36\Delta U}{5\hbar \omega_0}
\simeq 11.25 (\teps_{\rm cl}-\teps)^{5/4}.
\label{eq:strongfit}
\end{equation}
This result, shown in Fig. \ref{fig:epsplot}
by a solid line, agrees with the
numerical result for the strong impurity strength,
and gives a good estimate even for the weak impurity.
The expression (\ref{eq:strongfit}) may be useful to analyze
experimental data.

In order to analyze experimental results, it is convenient to
draw the graph of $\tilde{B}(\teps)$.
The feature of impurity effects can be obtained by this analysis.
Concerning the highly-biased region, this plot is universal
in the sense that it does not depend on the bare impurity strength $\eps$.

\begin{figure}[tb]
\hfil
\epsfxsize=8cm
\epsfbox{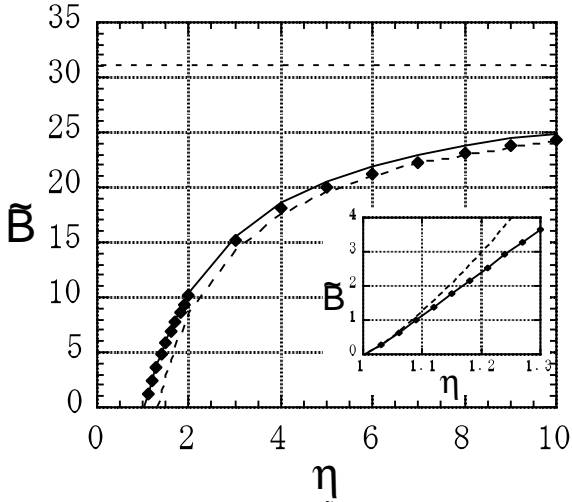}
\hfil
\caption{The scaled exponent $\tilde{B}$ versus 
the bias parameter $\eta = (f_{\rm c,0}-f)/(f_{\rm c,0}-f_c)$.
The numerical results are shown by square dots.
Dashed and solid curves correspond to the
analytical result for the weak impurity (\ref{eq:weakexp})
and for the strong impurity (\ref{eq:strongfit}),
respectively. The horizontal
dotted line shows the homogeneous case $\eps = 0$.
The solid and dashed curves in the inset 
show the behavior of $\tilde{B}$ 
near the threshold bias $\eta=1$, and
the asymptotic form (\ref{eq:strongfit2}) around $\eta=1$,
respectively.}
\label{fig:etaplot}
\end{figure}

In order to apply the viewpoint of the dimensional crossover,
the bias-dependence of the exponent $\tilde{B}$ is discussed.
For simplicity, the cosine potential with a bias term
(\ref{eq:pot1}) is considered with a notation
$f_{\rm c,0}=1$, (see above Eq. (\ref{homopot}))
$a=\sqrt{2}$, $b=1$ (see below Eq. (\ref{homopot})).
We use the bias parameter $\eta = (f_{\rm c,0}-f)/(f_{\rm c,0}-f_{\rm c})$.
It takes 0 and 1 corresponding the classical threshold ($f=f_{\rm c,0}=1$)
in the absence of the impurities and to the critical bias
($f=f_c$) in the presence of the impurities, respectively.
Here, $f_{\rm c}$ is defined by
$\teps_{\rm cl} =4/\sqrt{3} = \eps (2(1 - f_{\rm c}))^{-3/4}$.
The graph of $\tilde{B}$ as a function of $\eta$ is
shown in Fig. \ref{fig:etaplot}. The numerical result is shown by
square dots. Analytical results for
the weak and strong impurities are also shown by the dashed and 
solid curves, respectively.

The dimensional crossover discussed in Sec. \ref{sec:dimcross}
can be clarified in this figure.
In the region $\eta \gg 1$ where the effective impurity strength
$\teps$ becomes small, $\tilde{B}$ becomes almost 
independent of the bias as seen in Fig. \ref{fig:etaplot},
and the exponent $B$ behaves as $B\propto (f_{\rm c,0} - f)^1$
as expected in (\ref{eq:cs1}).
As $\eta$ approaches the critical threshold $\eta=1$, 
the effective impurity strength is enhanced,
and the exponent is suppressed. 
The analytical expression near $\eta=1$ is obtained from
(\ref{eq:strongfit}) as
\begin{equation}
B \simeq 22.36 (\eta-1)^{5/4}.
\label{eq:strongfit2}
\end{equation}
Hence, the exponent behaves as $B \propto (f_c-f)^{5/4}$
as expected in (\ref{exp1}).
We show the bias-dependences near $\eta=1$ 
in the inset of Fig. \ref{fig:etaplot}. 
The exponent, thus, behaves as 
$B \propto (f_c-f)^{3/2-(d+1)/4}$, and the spatial
dimension $d$ seems to be reduced to zero by the strong
impurities accompanied with the change of the classical 
threshold $f_c$.

\begin{figure}[tb]
\hfil
\epsfxsize=5.5cm
\epsfbox{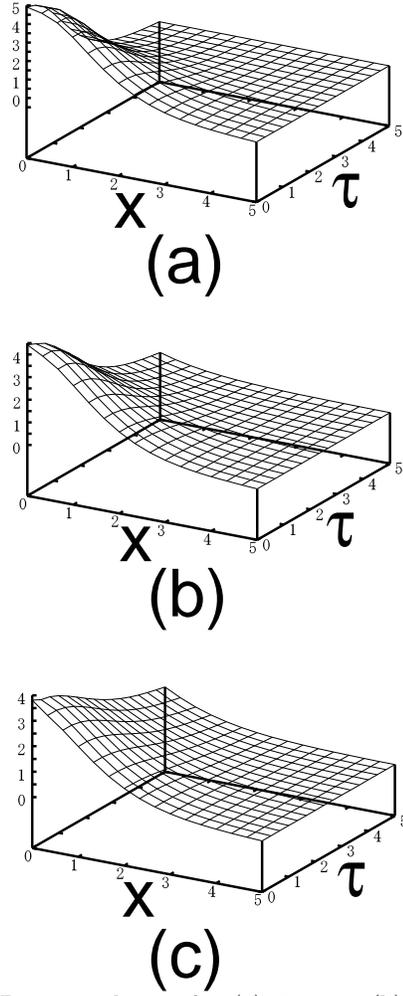}
\hfil
\caption{Bounce solution for (a) $\teps = 0$, (b) $\teps = 1.24$,
(c) $\teps = 1.98$. The critical value at the threshold bias
is $\teps = 4/\sqrt{3}=2.31$.}
\label{fig:bounce}
\end{figure}

Finally, we show the bounce solution $\vphi_{\rm B}(x,\tau)$
in Fig. \ref{fig:bounce} in the region $x>0$, $\tau>0$.
Note that the bounce solution has a reversal symmetry
$\vphi_{\rm B}(x,\tau)=\vphi_{\rm B}(-x,\tau)=\vphi_{\rm B}(x,-\tau)$.
The bounce solution is expected to describe the feature of the
nucleation process; $\vphi(x,\tau)$ describes the shape of the
field variable $\vphi$ at the imaginary-time $\tau$. 
As the effective impurity strength
increases, the boundary condition for the bounce solution $\vphi_{\rm B}
(x,\pm \infty)=\vphi_0(x)$ is modified at the impurity site. 
As seen in Fig. \ref{fig:bounce}, the nucleation process 
occurs only at the small region 
near the impurity site for the strong impurities. 
This change of the nucleation process
is responsible for the qualitative difference
between the weak and strong impurities.

\subsection{Prefactor}
\label{sec:prefactor}

The prefactor $A$ is ordinarily a minor factor which 
is hard to be measured experimentally.
However, it describes the system-size ($L$) 
dependence of the 
nucleation rate, which may be observed.
Here, we study the prefactor $A$ analytically
in the limiting cases within the bounce method, and
discuss the $L$-dependence of the nucleation rate.

The prefactor is formulated in (\ref{eq:prefactor})
with the improvement (\ref{eq:improvement})
due to the zero mode ($\lambda^{(B)}_2 = 0$).
In the absence of the impurities, not only the
second but also the third lowest modes
become the zero modes ($\lambda^{(B)}_3 = 0$),
which need to be treated by the Fadeev-Popov
method as
\begin{equation}
\frac{1}{\sqrt{\lambda^{\rm (B)}_3}} \rightarrow
\sqrt{\frac{\sigma \tilde{B}(0)}{2\pi}}
\int_{-\tilde{L}/2}^{\tilde{L}/2} dx_0' ,
\label{eq:improve2}
\end{equation}
where $x_0'$ denotes a position of the bounce center.
This treatment gives the correct system-size dependence,
$\Gamma \propto L$. 

As for the weak impurity,
the frequency of the third lowest mode is lifted to a small 
positive frequency. Also in this case, the replacement
in (\ref{eq:improve2}) is needed in order to 
obtain the correct system-size dependence. 
As a result, the tunneling rate is calculated as
\begin{equation}
\Gamma = \Gamma_0 \left[ \tilde{L} + \int_{-\tilde{L}/2}
^{\tilde{L}/2} dx_0' (e^{\sigma \teps f(x_0')} - 1) \right],
\end{equation}
where $\Gamma_0$ is the nucleation rate per unit length
in the absence of the impurities, and the function $f(x_0')$
is defined by
\begin{equation}
f(x_0') = \int_{-\infty}^{\infty} d\tau'
\vphi_{\rm B}(x_0',\tau';\teps = 0).
\end{equation}
From this expression, the tunneling rate can be divided into
two parts as
\begin{eqnarray}
\Gamma_{\rm bulk} &=& \Gamma_0 \tilde{L}, \\
\Gamma_{\rm imp} &=& \Gamma_0 \int_{-\tilde{L}/2}
^{\tilde{L}/2} dx_0' (e^{\sigma \teps f(x_0')} - 1).
\label{eq:gimp}
\end{eqnarray}
The bulk part $\Gamma_{\rm bulk}$ is proportional to the system size
$\tilde{L}$, while the impurity part $\Gamma_{\rm imp}$
is independent of the system size for $\tilde{L} \gg 1$.

\begin{figure}[tb]
\hfil
\epsfxsize=8cm
\epsfbox{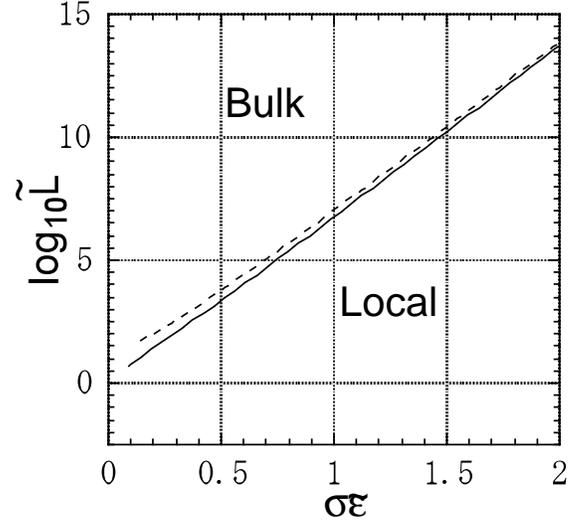}
\hfil
\caption{Crossover length $\tilde{L}_{\rm cr}=L_{\rm cr} a^{1/2}
(f_c-f)^{1/4}$ versus $\teps \sigma = \eps a^{1/2}
(f_c-f)^{1/4}/b$.
In the region $\tilde{L}>\tilde{L}_{\rm cr}$, 
nucleation occurs in the bulk region,
while in the region $\tilde{L}<\tilde{L}_{\rm cr}$,
the impurity contribution $\Gamma_{\rm imp}$ becomes dominant. 
The solid and dashed lines show the result and,
the asymptotic expression (\ref{eq:crossasy}) 
valid in the limit $\sigma \teps \gg 1$, respectively.}
\label{fig:crossover}
\end{figure}

Within the bounce method, the $L$-dependence is
naturally described as follows.
For large $L$, $\Gamma_{\rm bulk}$
alway dominates $\Gamma_{\rm imp}$,
and no impurity effects appear.
As $L$ is reduced, the bulk part is suppressed,
and below a crossover length $\tilde{L}_{\rm cr}$, 
the impurity part may overcome the bulk part.
Only in this region, the impurity effect is clearly
observed experimentally.
The ratio $\Gamma_{\rm imp}/\Gamma_{\rm bulk}$ is determined
by the scaled length $\tilde{L}$ and the impurity factor
$\sigma \teps$ in (\ref{eq:gimp}).
For $\tilde{L} \gg 1$, the crossover length $\tilde{L}_{\rm cr}$ 
can be obtained as a function of $\sigma \teps$
by taking the ratio as $\Gamma_{\rm imp}/\Gamma_{\rm bulk}=1$.
Thus, the phase diagram is obtained by Fig. \ref{fig:crossover},
where the crossover length is shown by the solid curve. 
In the upper region of the phase diagram, 
the nucleation occurs dominantly in the
bulk region, while in the lower region, the impurity part dominates 
the bulk part, and inhomogeneity effects appear clearly.

The intuitive discussion about the appearance of
the impurity effect in Sec. \ref{sec:model} can be now
clarified in detail. 
The asymptotic form of the crossover length $\tilde{L}_{\rm cr}$ 
for $\sigma \teps \gg 1$ is obtained analytically 
by applying the stationary method to the integral (\ref{eq:gimp}) as
\begin{equation}
\log \tilde{L}_{\rm cr} \simeq 16.43 (\sigma \teps) - 0.211
- \frac{1}{2} \log (\sigma \teps).
\label{eq:crossasy}
\end{equation}
This result is shown in Fig. \ref{fig:crossover} by the dashed curve.
The predominant term in (\ref{eq:crossasy}) can be related
to the enhancement of the exponent $\Delta B$ defined by
\begin{equation}
\Delta B = \sigma \teps \int_{-\infty}^{\infty}
d\tau' \vphi_{\rm B}(x=0,\tau';\teps=0),
\end{equation}
as $\log \tilde{L}_{\rm cr}=\Delta B$.
Since $\Delta B$ is estimated by the local suppression of
the potential barrier $\Delta U$ and the typical attempt frequency
$\omega_0$ as $\Delta B \sim \Delta U/\hbar \omega_0$,
the expression in (\ref{condition}) can be reproduced
qualitatively.

Beyond the perturbational analysis of $\teps$, it is difficult to
calculate the prefactor, because the low-frequency modes
may couple with each other. It would be a future problem
to consider the nucleation rates in this regime. (See also
Sec. \ref{sec:validity}.) Only for the strong impurity, 
the prefactor can be calculated by assuming $\tilde{L}
\ll \tilde{L}_{\rm cr}$ and by applying standard procedures 
\cite{Coleman77,Coleman79,Weiss93}
to one-variable problem (\ref{aaa1}) with (\ref{aaa2}) as
\begin{equation}
A \sim \sqrt{60} \sqrt{\frac{\tilde{B}(0)}{2\pi}},
\end{equation}
although the crossover length $\tilde{L}_{\rm cr}$
cannot be easily obtained even in this case.

\section{THERMAL ACTIVATION REGIME}
\label{sec:thermal}

At high temperatures, 
nucleation is caused by thermal fluctuations.
Since the nucleation rate has already 
been studied in this region in ref. \cite{Kato00}, 
we only summarize the feature of the impurity effect.

In the thermal activation regime, 
the nucleation rate is evaluated by Kramers' formula
\begin{equation}
\Gamma = A \exp(-\Delta U/k_{\rm B}T).
\end{equation}
The potential barrier $\Delta U$ is calculated as
\begin{eqnarray}
\Delta U &=& U[\vphi_B(x')] - U[\vphi_0(x')], \\
U[\vphi(x')] &=& \frac{a^{5/2}(f_c-f)^{5/4}}{b^2}
\int_{-\tilde{L}/2}^{\tilde{L}/2} dx' \nonumber \\
&\times& \left( \frac12 \vphi_{x'}^2 + \frac12 \vphi^2
-\frac16 \vphi^3 - \teps \delta(x') \vphi(x') \right).
\end{eqnarray}
Here, $\vphi_0(x')$ ($\vphi_{\rm B}(x')$) is the stable
(unstable) stationary solution satisfying
\begin{equation}
\frac{\delta U}{\delta \vphi(x')}=
-\vphi_{x'x'}+\vphi-\frac{\vphi^2}{2}-\teps \delta(x') = 0.
\end{equation}
The same result can be derived by applying
the bounce method to a finite temperature region
\cite{Langer67,Langer69,Affleck81,Grabert84,Larkin84}.
Within the bounce method, the feature at high temperatures
is explained as follows. 
At high temperatures, the bounce solution becomes
independent of $\tau'$, since the interval
in the $\tau$ direction becomes short.
It means that the temporal dimension become irrelevant. 
Hence, the relevant dimension of the bounce is reduced 
from $(d+1)$ to $d$. Except for the absence of the
temporal dimension, we can discuss the dimensional crossover
in the same way as Sec. \ref{sec:dimcross}:
the energy barrier behaves as $\Delta U
 \propto (f_c - f)^{5/4}$
for the weak impurity, while as $\Delta U \propto (f_c -f)^{3/2}$
for the strong impurity.

\begin{figure}[tb]
\hfil
\epsfxsize=8cm
\epsfbox{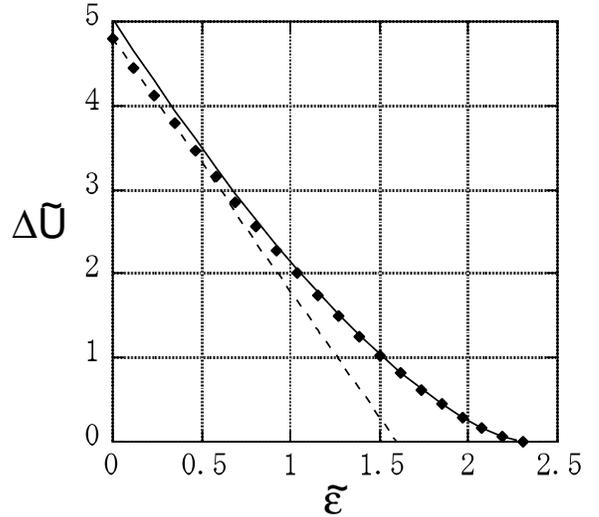}
\hfil
\caption{The normalized barrier height $\Delta \tilde{U}$
versus the effective impurity strength $\teps$.
The numerical results are shown by the square dots, while
analytical results in the limiting case, (\ref{eq:strongimp22}) 
and (\ref{eq:strongimp2}), are shown by the dashed and solid curves,
respectively.}
\label{fig:thermal}
\end{figure}

The barrier height is calculated in a form
\begin{equation}
\Delta U = \frac{a^{5/2}(f_c-f)^{5/4}}{b^2} \Delta
\tilde{U} (\teps).
\end{equation}
The impurity effects appear only through the normalized potential
barrier $\Delta \tilde{U}$. We show $\Delta \tilde{U}$ as a function of the
effective impurity strength $\teps$ in Fig. \ref{fig:thermal}
by the square dots.
The critical impurity strength is given by 
$\teps_{\rm cl} = 4/\sqrt{3}$ 
which is the same as the quantum tunneling regime.
The exponent is obtained analytically for the weak impurity 
$\teps \ll 1$ as
\begin{equation}
\Delta \tilde{U}(\teps) = 24/5 - 3\teps,
\label{eq:strongimp22}
\end{equation}
while for the strong impurity $\teps \sim \teps_{\rm cl}$ as
\begin{equation}
\Delta \tilde{U}(\teps) = 1.433 (\teps_{\rm cl}-\teps)^{3/2}. 
\label{eq:strongimp2}
\end{equation}
These results are shown in Fig. \ref{fig:thermal} by
the dashed and solid curves. The feature of the impurity effect is
the same as the zero-temperature case shown in Fig. \ref{fig:epsplot}.

\section{DISCUSSION}
\label{sec:discussion}

\subsection{Extension to the $d \ge 2$ case}

First, let us discuss the two-dimensional systems. 
The stable solution $\vphi_0(x',y')$ is determined by 
\begin{equation}
\vphi_{x'x'} + \vphi_{y'y'} 
= \vphi - \frac12 \vphi^2 - \teps \delta (x') \delta(y'),
\end{equation}
where the last term describes the impurity located at
$(x',y')=0$. For the stationary solution having
a rotational symmetry around
the origin $(x',y')=(0,0)$, this equation is reduced,
by using the radius coordinate $r=\sqrt{x'^2+y'^2}$, to
\begin{equation}
\frac{1}{r} \frac{\partial}{\partial r}\left( r
\frac{\partial \vphi}{\partial r} \right) = \vphi - \frac12
\vphi^2 - \teps g(r).
\label{eq:bounce3D}
\end{equation}
Here, $g(r)$ is a modified delta function satisfying
$g(r)=0$ for $r>0$, $g(r) =\infty$ for $r=0$, and
\begin{equation}
\int_0^{\infty} dr 2\pi r g(r) = 1.
\label{deltag}
\end{equation}
From (\ref{eq:bounce3D}) and (\ref{deltag}), the boundary condition
at $r=0$ is obtained as
\begin{equation}
\lim_{r \rightarrow 0} r \frac{\partial \vphi}{\partial r}
= - \frac{\teps}{2\pi}.
\end{equation}
Since the function $r \vphi_r$ is a continuous function
at $r>0$, if we take an arbitrary constant $A$ ($< \teps/2\pi$),
there exists a constant $a$, and for all $r$ 
in the range $0 < r \le a$
\begin{equation}
r \frac{\partial \vphi}{\partial r} \le -A
\end{equation}
is satisfied. As a result, $\vphi(r)$ is evaluated for $0<r<a$ as
\begin{equation}
\vphi(r) \ge A \log \left(\frac{a}{r} \right) + \vphi(a),
\end{equation}
and the value of $\vphi$ diverges at $r=0$.
Since $\vphi$ must be finite in the present model, this
divergence means that the local
inhomogeneity cannot be described by the delta function
in the two-dimensional systems. 
In other words, if we approximate the delta function as
$g(r) = 1/\pi r_0^2$ for $0<r<r_0$ and as 
$g(r) = 0$ for $r_0<r$, the value of $\vphi(0)$
depends crucially on the impurity size $r_0$, and
diverges in the limit $r_0 \rightarrow 0$.
Also in the $d=3$ case, the field $\vphi(r)$ diverges at $r=0$, 
since the bounce solution behaves
$\vphi(r) \propto 1/r$ near $r=0$, where $r=\sqrt{x'^2+y'^2+z'^2}$,

These results indicate that in the highly-biased region,
the system is sensitive to local inhomogeneities.
To clarify the inhomogeneity effects in the $d \ge 2$ case,
it is needed to consider the detailed profiles of
the inhomogeneities beyond the local approximation 
adopted in this paper.

\subsection{Validity of the bounce method}
\label{sec:validity}

In this paper, the nucleation rate is calculated by the bounce method. 
Although this method has been applied to multi-variable tunneling
problems for many years, we should be careful of its validity.
In principle, the exact tunneling rate should be determined 
by the full information
about the action $S[\phi(x')]$. The bounce solution
can extract important informations from the action;
the exponent $B$ is a difference of the action between
the bounce and stable solutions, while
the prefactor $A$ is determined by the second-order fluctuations 
around both the solutions. While the bounce method is a 
convenient method to evaluate the tunneling rate, some informations which
may affect the tunneling rate are dropped.

It is expected that the bounce method can be justified in the 
following conditions \cite{Weiss93}: 
(1) we can define the tunneling variable 
describing slow dynamics among the many degrees of freedom,
and (2) this tunneling variable is well separated from the other
variables with fast dynamics.
In these conditions, the system can be reduced to the one-variable
tunneling problem where the bounce method is justified well.
Here, let us consider the model adopted in this paper.
For the strong impurity, it has been shown that
the single-mode approximation gives a good estimate of
the exponent $B$. In this region, the bounce method is
expected to give correct results for both the exponent and
the prefactor for $L\ll L_{\rm cr}$, though the crossover length
$L_{\rm cr}$ itself cannot be evaluated. 
For the moderate impurity, however, there 
appears another low-frequency mode related to the Goldstone mode. 
This low-frequency mode
is expected to couple with the tunneling mode. 
Hence, the bounce method may give
an incorrect result especially for the prefactor determined
only by the information about second-order fluctuations.
It is a future problem to improve the bounce method in this situation.

\section{SUMMARY}
\label{sec:summary}

The local inhomogeneity effects on nucleation processes have
been studied within the bounce method at high biases.
The nucleation rate has been calculated in a form $\Gamma =
A \exp (-B/\hbar)$. It has been found that the effective
impurity strength $\teps$ can be controlled not only by
the bare impurity strength $\eps$ but also by the external
bias $f$. The exponent $B$ has been calculated as a function of 
the effective impurity strength $\teps$
without reduction to a single-variable tunneling problem.
It has been shown that these results are well reproduced
for the weak impurity by a perturbational analysis and for the strong
impurity by a one-mode approximation.
It has been clarified that the results can be explained
by the reduction of the effective dimensionality of the system
due to the impurity.
By calculating the prefactor $A$,
the condition of the appearance of the impurity effects has
been discussed, and the phase diagram has been obtained.

There remain future problems. One of them is to study 
the inhomogeneity effect in higher-dimensional systems.
As discussed in Sec. \ref{sec:discussion},
the local approximation does not hold for the $d\ge 2$ case.
From this result, it is conjectured that the sample 
inhomogeneities may become essentially important for 
higher-dimensional systems.
I think that the study of the inhomogeneity effects in
high-dimensional systems would give an important aspect about
the interplay of quantum tunneling and randomness.

\acknowledgments

I wish to acknowledge M. Yumoto for suggestions and helpful discussions.

\appendix
\section{Relevance of the Local Impurity Model to Experimental Systems}
\label{app}

In this paper, the inhomogeneity effect is described by the
potential
\begin{equation}
V_{\rm imp}(x,\phi) = \eps(x) h(\phi).
\label{eq:Vimp3}
\end{equation}
If the function $\eps(x)$ changes only in a narrow region
at the impurity site compared with the typical length
scale ($\sim a^{-1/2} (f-f_c)^{-1/4}$), it can be replaced
with the delta function $\delta(x)$ in (\ref{Vimp}).
The potential form (\ref{eq:Vimp3}) can be realized 
in various experimental situations. One typical example is
a spatial inhomogeneity of the bias. If the bias is assumed
not to be uniform but to have a spatial dependence as $f + \eps(x)$,
then we have the potential form (\ref{eq:Vimp3}). Another example is
the inhomogeneity of the amplitude of the potential barrier
in $V_0(\phi)$. If we replace, for example, the cosine potential
in (\ref{eq:pot1}) with $(1+\eps(x))(1-\cos \phi)$, then the
form (\ref{eq:Vimp3}) is also obtained. As for the Josephson
junction systems, the coupling to the
derivative of the field $\phi_x(x)$ can also be introduced
by applying the magnetic field \cite{Ustinov99}. 
If the coupling is denoted with $b(x)\phi_x(x)$, the potential
form (\ref{eq:Vimp3}) is derived by taking the function $\eps(x)$
as $-b_x(x)$.

In addition to the above situations, boundaries play
a role of local inhomogeneities. The potential barrier
$\Delta U$ is reduced to $\Delta U/2$ at open
edge. Hence, the nucleation occurs dominantly at the boundary.
The prefactor $A$ of the nucleation rate
is always independent of the system size, while
the exponent $B$ is affected by the boundary condition.
Assuming that the field $\vphi(x')$ is
defined at $x' \ge 0$, and that the boundary condition is given as
\begin{equation}
\vphi_{x'} (x'=0) = h,
\end{equation}
it can be shown easily that this boundary effect corresponds to
the local inhomogeneity by taking $\teps = 2 h$. Thus, the exponent
$B_{\rm edge}$ is obtained as
\begin{equation}
B_{\rm edge} = \frac12 \sigma \tilde{B}(\teps=2h).
\end{equation}
By changing the boundary condition
at the edge, the impurity effects on the exponent
can also be observed. Actually, this tendency
has been observed in experiments \cite{Castellano96}.


\begin{thebibliography}{99}
\bibitem{Haenggi90}P. H\"anngi, P. Talkner, and M. Borkovec,
Rev. Mod. Phys. {\bf 62}, 251 (1990).
\bibitem{Lifshitz72}I. M. Lifshitz, and Yu. Kagan,
Zh. Eksp. Teor. Fiz. {\bf 62}, 385 (1972) 
[Sov. Phys. JETP {\bf 35}, 206 (1972)].
\bibitem{Satoh91}T. Satoh, M. Morishita, M. Ogata, A. Savada, 
and T. Kuroda, Physica B {\bf 169}, 513 (1991).
\bibitem{Satoh92}T. Satoh, M. Morishita, M. Ogata and S. Katoh,
Phys. Rev. Lett. {\bf 69}, 335 (1992).
\bibitem{Balibar95}S. Balibar, C. Guthmann, H. Lambare, P. Roche,
E. Rolley, and H.J. Maris, J. Low Temp. Phys. {\bf 101}, 271 (1995).
\bibitem{Ruutu96}J. P. Ruutu, P. J. Hakonen, J. S. Penttil\"a,
A. V. Bobkin, J. P. Saram\"aki, and E. B. Sonin,
Phys. Rev. B {\bf 77}, 2514 (1996).
\bibitem{Caldeira88}A. O. Caldeira, and K. Furuya, J. Phys. C,
{\bf 21}, 1227 (1988).
\bibitem{Chudnovsky88}E. M. Chudnovsky, and L. Gunther, Phys. Rev. B
{\bf 37} 9455 (1988).
\bibitem{Chudnovsky98}E. M. Chudnovsky, and J. Tejada,
{\it Macroscopic Quantum Tunneling of the Magnetic Moment}
(Cambridge University Press, Cambridge, 1998).
\bibitem{Takeuchi00}S. Takeuchi, T. Suzuki, and H. Koizumi,
J. Phys. Soc. Jpn. {\bf 69}, 1727 (2000).
\bibitem{Petukhov98}B. V. Petukhov, H. Koizumi, and T. Suzuki,
Philos. Mag. A {\bf 77}, 1041 (1998).
\bibitem{Maki78}K. Maki, Phys. Rev. B {\bf 18}, 1641 (1978).
\bibitem{Hida84}K. Hida, U. Eckern, Phys. Rev. B {\bf 30},
4096, (1984).
\bibitem{Nakaya86}S. Nakaya, and K. Hida, J. Phys. Soc. Jpn.
{\bf 55}, 3768 (1986).
\bibitem{Duan93}Ji-Min Duan, Phys. Rev. B {\bf 48}, 4860 (1993).
\bibitem{Maki95}K. Maki, Phys. Lett. A {\bf 202}, 313 (1995)
\bibitem{Hatakenaka98}N. Hatakenaka, M. Shiobara, Ken-ichi Matsuda,
and S. Tanda, Phys. Rev. B {\bf 57}, 2003 (1998).
\bibitem{Zotov93}S. V. Zaitsev-Zotov, Phys. Rev. Lett. {\bf 71}, 605
(1993).
\bibitem{Zotov94}S. V. Zaitsev-Zotov, Phys. Rev. 
Lett. {\bf 72}, 587 (1994).
\bibitem{Zotov97}S. V. Zaitsev-Zotov, G. Remenyi, and P. Monceau, 
Phys. Rev. B 56, 6388 (1997).
\bibitem{Davidson85}A. Davidson, B. Dueholm, B. Kryger,
and N. F. Pedersen, Phys. Rev. Lett. {\bf 55}, 2059 (1985).
\bibitem{Ustinov92}A. V. Ustinov, T. Doderer, R. P. Huebener,
N. F. Pedersen, B. Mayer, and V. A. Oboznov, Phys. Rev. Lett.
{\bf 69}, 1815 (1992).
\bibitem{Castellano96}M. G. Castellano, G. Torrioli, C. Cosmelli,
A. Costantini, F. Chiarello, P. Carelli, G. Rotoli, M. Cirillo,
and R. L. Kautz, Phys. Rev. B {\bf 54}, 15417 (1996).
\bibitem{Kato96}T. Kato, and M. Imada, J. Phys. Soc. Jpn.
{\bf 65}, 2963 (1996).
\bibitem{Ustinov99}A. V. Ustinov, B. A. Malomed, and E. Goldobin,
Phys. Rev. B {\bf 60}, 1365 (1999).
\bibitem{effective}For example, see ref. \cite{Ruutu96}.
\bibitem{Coleman77}S. Coleman, Phys. Rev. D {\bf 15}, 2929 (1977). 
\bibitem{Callan77}C. G. Callan, and S. Coleman, Phys. Rev. D {\bf 16},
1762 (1977).
\bibitem{Coleman79}S. Coleman, {\it The Use of Instanton} in 
{\it The Whys of Subnuclear Physics}, edited by A. Zichichi (Plenum, New
York), p. 805.
\bibitem{Kramers40}H. A. Kramers, Physica (Utrecht) {\bf 7}, 284 (1940).
\bibitem{Yumoto99}M. Yumoto, H. Fukuyama, H. Matsukawa, and N. Nagaosa,
J. Phys. Soc. Jpn. {\bf 68}, 170 (1999).
\bibitem{Yumoto00a}M. Yumoto, H. Fukuyama, H. Matsukawa, and N. Nagaosa,
Physica B {\bf 284-288}, 1667 (2000).
\bibitem{Yumoto00b}M. Yumoto, H. Fukuyama, H. Matsukawa, and N. Nagaosa,
J. Phys. Soc. Jpn. {\bf 69}, 2953 (2000).
\bibitem{Kato00}T. Kato, J. Phys. Soc. Jpn. {\bf 69}, 2735 (2000).
\bibitem{McLaughlin78}D. W. McLaughlin, and A. C. Scott,
Phys. Rev. A {\bf 18}, 1652 (1978).
\bibitem{Kivshar89}Y. S. Kivshar and B. A. Malomed, Rev. Mod. Phys.
{\bf 61}, 763 (1989).
\bibitem{Ivlev87}B. I. Ivlev, and V. I. Mel'nikov, Phys. Rev.
B {\bf 36}, 6889 (1987).
\bibitem{Weiss93}U. Weiss, {\it Quantum Dissipative Systems}
(World Scientific, Singapore, 1993).
\bibitem{Langer67}J. S. Langer, Ann. Phys. {\bf 41}, 108 (1967).
\bibitem{Langer69}J. S. Langer, Ann. Phys. {\bf 54}, 258 (1969).
\bibitem{Affleck81}I. Affleck, Phys. Rev. Lett. {\bf 46}, 388 (1981).
\bibitem{Grabert84}H. Grabert and U. Weiss,
Phys. Rev. Lett. {\bf 53}, 1787 (1984).
\bibitem{Larkin84}A. I. Larkin and Yu. N. Ovchinnikov, Zh. Eksp.
Theor. Fiz. {\bf 86}, 719 (1984) [Sov. Phys.-JETP {\bf 59}, 420 (1984)].
\end{thebibliography}
\end{document}